# Predicting pathways for old and new metabolites through clustering

Thiru Siddharth[1], Nathan E. Lewis[2]

[1] Department of Computer science and Engineering, Indian Institute of Information Technology, Bhopal, MP 462003, INDIA
[2] Department of Pediatrics and Bioengineering, University of California San Diego, La Jolla, CA 92093, USA

Abstract: The diverse metabolic pathways are fundamental to all living organisms, as they harvest energy, synthesize biomass components, produce molecules to interact with the microenvironment, and neutralize toxins. While discovery of new metabolites and pathways continues, the prediction of pathways for new metabolites can be challenging. It can take vast amounts of time to elucidate pathways for new metabolites; thus, according to HMDB only 60% of metabolites get assigned to pathways. Here, we present an approach to identify pathways based on metabolite structure. We extracted 201 features from SMILES annotations, and identified new metabolites from PubMed abstracts and HMDB. After applying clustering algorithms to both groups of features, we quantified correlations between metabolites, and found the clusters accurately linked 92% of known metabolites to their respective pathways. Thus, this approach could be valuable for predicting metabolic pathways for new metabolites.

## 1. INTRODUCTION

Every living organism leverages diverse metabolic pathways to support a wide range of biological functions. The study of these pathways helps researchers design new drugs, unravel disease mechanisms, and engineer cells with valuable metabolic and biosynthetic capabilities. These pathways have many constituents like enzymes, catalysts and different types of chemical compounds involved in the reactions. Every year many new metabolites are found but the biosynthetic and catabolic pathways remain unknown for many. Pathways prediction methods are invaluable to researchers as they aim to link metabolites to metabolic pathways.

   The use of machine learning to reconstruct metabolic pathways is of great interest given its remarkable capacity for handling large and complicated datasets being gathered through numerous initiatives. Many studies stand to benefit from such efforts, given the recent development of tools for modeling and reconstructing metabolic pathways. Some computational techniques for the creation and reconstruction of metabolic pathways have been reviewed by Wang et al. [1]. Cuperlovic-Culf [2] also evaluated relevant literature on the use of machine learning to simulate metabolic pathways. Kim et al. [3] have compiled the machine learning algorithms in systems metabolic engineering.

   An approach for predicting metabolic pathways from an organism's annotated genome has been previously proposed [4]. Metabolic pathways are among the subsystems that the SEED [5,6] projects into genomes. Proposed subsystems are computationally inferred and then verified and improved by curators. Predictions must be verified, and a variety of approaches can be used [7]. Reactome [8] has also been used to predict the metabolic pathway for 5-aminoimidazole ribonucleotide biosynthesis II using log-odds ratios and feature values for a naive Bayes predictor built with HC-AIC feature selection and trained on the complete gold standard. However, more evidence is needed to validate the accuracy of the approach. KEGG [9] can also be used to generate "pathway maps" using genomic data. Because KEGG pathway maps include numerous metabolic pathways from different species, KEGG confronts the route map prediction difficulty as opposed to the pathway prediction problem [10]. Other algorithms have been proposed to predict metabolic pathways using constraint-based modeling. Phylogenetic reconstruction and the connection between metabolic pathways and phenotypes have both been accomplished using extensive reconstructions of metabolites at the level of the pathway [11,12,13]. The existence or absence of pathways has often been determined by these attempts using straightforward rules or scores. A route is usually assumed to require enzymes for all reaction steps to be deemed likely [11], unless some reactions are thermodynamically deemed to be spontaneous. In their score, Kastenmüller et al. [12,13] Computed the proportion of reactions existing in the pathway, weighting each reaction according to its individuality. This score compares to the "information content" characteristics employed by the predictors. Using probabilities of pathway existence calculated by these and other approaches could help identify pathways. The reactome prediction problem has been approached by several studies and methods have been proposed, such as IdentiCS [14], metaSHARK [15], and Pathway Analyst [16,17], wherein they assign enzymes to the reactions they catalyze using various sequence analysis techniques. It remains unclear, however, how enzyme/reaction mappings are best used to determine whether a pathway is present in an organism. A route is only deemed present by Pathway Analyst if at least one of its reactions contains an enzyme. However, pathways may involve many enzymatic steps, and it is important to prioritize predictions based on enzymatic and/or thermodynamic support. Numerous methods have been proposed for discovering (or developing) new pathways, including search-based techniques (like those in [18]), which identify potential pathways between both input and output molecules. Another strategies involve looking for sequences of molecular functionalities in biological networks [19] or using kernel-based techniques to discover correlations between the enzymes that catalyze consecutive reactions in metabolic pathways [20]. These techniques offer helpful tools for locating novel pathways that may be tested experimentally and added to curated pathway databases like MetaCyc. In this regard, the pathway prediction techniques we have discussed here are usefully supplemented by pathway discovery techniques.

In this study we present an alternative approach to classify pathways for metabolites using clustering techniques, including Kmode clustering and k-prototyping clustering. We have tested our model on metabolites extracted from PubMed, HMDB and demonstrated that it successfully links known metabolites to their associated pathways. Thus, this method could be applied to annotate new metabolites and guide further experiments identifying and characterizing enzymes associated with new metabolites.

## 2. METHODS

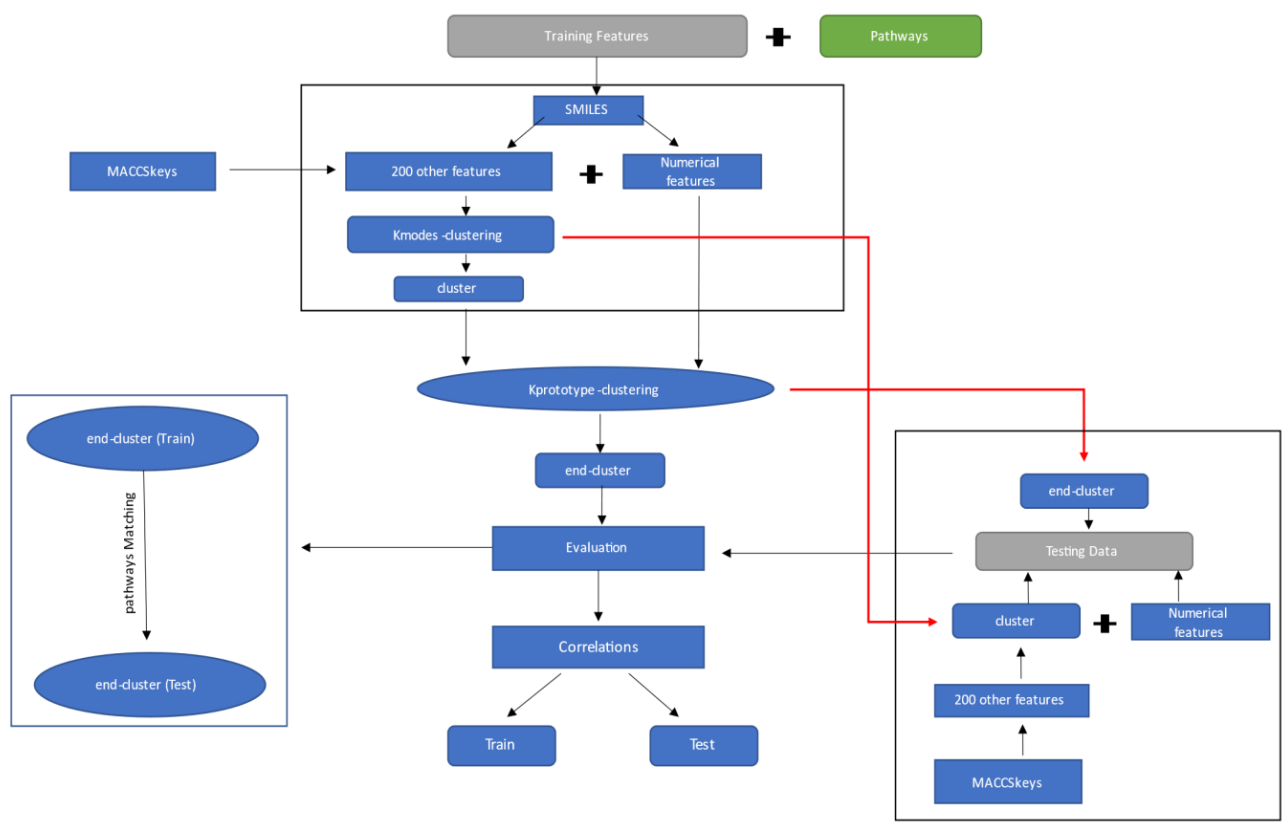

*Figure 1* **Block Diagram of the method.**

2.1 Dataset

To enable model training, data were extracted from HMDB and the model was tested on the data extracted from PubMed. This dataset was used to evaluate the pathways including new pathways and old pathways. The dataset consists of two groups of features. The first group contains MACCSkeys of metabolites and the other includes physical properties of metabolites. MACCSKeys (fingerprints) were generated with RDKit and physical properties were generated using the online web tool cactus (PubChem) using SMILES (Simplified Molecular Input Line Entry System) of molecules. 167 features were generated from a single MACCSkey and another 34 features are physical features. Around 3500 metabolites were extracted from HMDB for training and 2000 were extracted from PubMed for testing. Later pathways were extracted using HMDB IDs using the 'Beautifulsoup' python library for the respective known metabolites. After that, both groups of features integrated with each other, in which some features only had a single type of entity or value, like "UndefinedBondStereoCount" and "IsotopeAtomCount".

## 2.2 Pre-processing

Before deploying the two types of clustering, kmodes clustering and, K-prototype clustering, data pre-processing was done by using Principal component Analysis (PCA). Moreover, machine learning algorithms were also applied to evaluate all accuracies.

PCA was applied for reducing the dimensionality of the data. In **Figure 2** PCA scatter plots are shown for training and testing features.

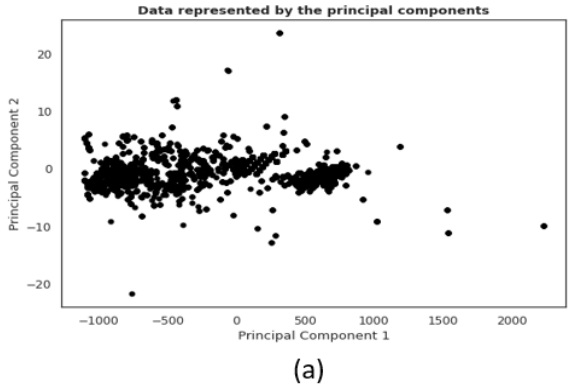
(a)

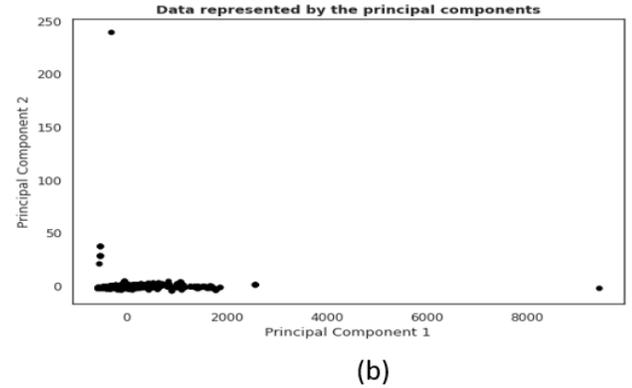
(b)

*Figure 2 PCA scatter plots (a) and (b) are given for training and testing datasets respectively*

### 2.3 Clustering

2.3.1 Kmodes clustering

Categorical data cannot be clustered using K-means clustering [21,22]. Thus, we used K-modes clustering, which adapts the paradigm of k-means to cluster categorical variables by employing a basic fitting of distinct data objects for categorical elements, modes rather than cluster averages, and a frequency-based mechanism to modify modes in the k-means way to reduce the clustering cost function. Since the k-modes methodology employs a similar clustering method as the k-means approach, its efficacy is preserved [20]. Because of the many measures it is using, k-modes clustering does not delete the data from the K-means pattern; it restricts the use of numbers while maintaining efficacy.

.
2.3.2 K-prototype clustering

To handle clustering methods with heterogeneous data types [23], Huang introduced an algorithm called K-Prototype (numerical and categorical variables). A clustering technique based on partitioning is K-Prototype. To handle clustering with heterogeneous data types, its algorithm is an upgrade of the K-Means and K-Mode clustering algorithm. The k-means and k-modes algorithms can easily be incorporated into the k-prototypes algorithm that is used to group the objects of mixed types [24,25].

The goal of the k-prototype was to divide the metabolite's dataset into 'k' clusters by solving the optimization problem[26].

$$E = \sum_{l=1}^{k} \sum_{i=1}^{n} u_{il} d(x_i, Q_1) \quad \ldots \quad (3.1)$$

Where,
$U_{il}$ = Element of partition matrix $U_{n*k}$.
$Q_i$ = prototype or cluster $l$ *vector*.

$d(x_i, Q_i)$ = measure of dissimilarity defined as :

$$d(x_i, Q_l) = \sum_{j=1}^{p} \left(x_{ij}^r - q_{lj}^r\right)^2 + \mu_l \sum_{j=p+1}^{m} \delta\left(x_{ij}^c, q_{lj}^c\right) \ldots \quad (3.2)$$

Where, $\delta(p, q) = 1$ if $p \neq q$
$\delta(p, q) = 0$ if $p = q$
$x^r_{ij} (x^c_{ij})$ = value of the jth categorical attribute in the $l$ cluster.
$\mu_l$ = weight for categorical attributes in the $l$ cluster.

The algorithm was implemented with two objectives in mind: first, to retain the uncertainties associated with data sets for a longer time before conclusions are drawn; and second, to consider the significance of specific variables to the clustering process. Created metabolite's dataset had categorical and numeric type of features. That is why K-prototype clustering was used to get clusters from numeric and categorical features of metabolite's dataset.

The leveraged k-prototype algorithm's operations are explained as follows:

Step-1: From the metabolite's dataset for k clusters, algorithm picks k initial prototypes.
Step-2: Each data item in the dataset gets assigned to the cluster with the closest prototype. In Eq. (3.2) After each allocation, cluster prototype gets updated.
Step-3: After each data object has been assigned to a cluster, re-evaluation of how closely the new data objects resemble the old prototypes is done. when it turns out that a data item's closest prototype belongs to a different cluster than the one it was in, reallocate it to that cluster and update the prototypes for both clusters.

Step-4: Continue Step 3 until no data object has changed clusters after a complete cycle test of dataset.

## 2.4 Performance measure

Multiple ML algorithms were applied to the data, including Extra Trees Classifier, Ada Boost Classifier, Gradient Boosting Classifier, Light Gradient Boosting Machine, Quadratic Discriminant Analysis, Random Forest Classifier, SVM - Linear Kernel, K Neighbors Classifier, Logistic Regression, Decision Tree Classifier, Light Gradient Boosting Machine, Naive Bayes, Ridge Classifier, Linear Discriminant Analysis and Dummy Classifier. Also, several metrics were computed for each algorithm as shown in **Table 1**, such as accuracy, f1-score, recall, precision score, AUC, ROC, and Cohen's Kappa score. The highest accuracy achieved by the Adaboost classifier with auc of 92% and f1 of 91% respectively.

### 2.4.1 Evaluation Metrics

$$Specificity = \frac{|TP|}{|TP|+|FN|} \quad (3.3)$$

$$Sensitivity = \frac{|TP|}{|TP|+|FN|} \quad (3.4)$$

$$F1 = \frac{2 \times precision \times recall}{precision + recall} \quad (3.5)$$

The first of two additional coefficients in **Table 1** is the Cohen coefficient. The Cohen's Kappa (k) metric has some recognition. Measures of our model's performance in multi-class classification problems, such as precision, recall, and accuracy, may not necessarily provide a clear picture. We could also run into the problem of unbalanced classes. Think of two categories, Z and Y, for instance, where Z only appears 10% of the time. We rely on metrics like recall and precision since accuracy can be deceptive. Furthermore, we may combine these metrics to create a single measurement called the F1 score, recall measurements and the harmonic mean of the accuracy.

Equation (3.6) provides the Cohen's kappa statistic, which addresses both multi-class and unbalanced class difficulties.

$$k = \frac{p_o - p_e}{1 - p_e} \quad (3.6)$$

Where $p_o$ represents the observed agreement and $p_e$ the expected agreement. It provides details on the classier performance. Cohen's kappa can never be greater than one and is normally one or less. If the classifier returns a value of 0 or less, it is ineffective. Its values cannot all be interpreted in the same way.

Second is the Matthews correlation coefficient (MCC) is given using Equation (3.7). It is a measure of the validity of binary classifications.

$$MCC = \frac{(TP \times TN) - (FP \times FN)}{\sqrt{(TP+FP)(TP+FN)(TN+FP)(TN+FN)}} \quad (3.7)$$

### 2.4.2 AdaBoost

Boosting techniques can combine several low accuracy models to create higher accuracy models. AdaBoost was tested, Gradient Boosting, and XGBoost were tested. When both features, end-cluster and cluster were fed in multiple classification algorithms, after evaluation of all techniques, it is concluded that AdaBoost achieved the highest accuracy as shown in **Figure 3**.

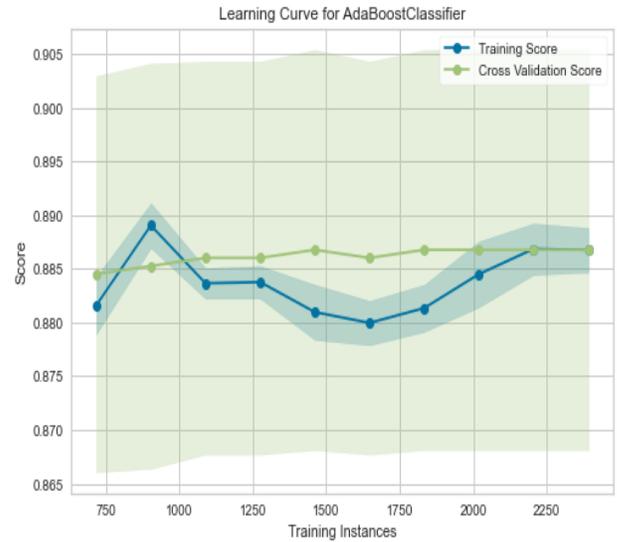

*Figure 3 Training and Cross validation score plot for AdaBoost*

$$H(x) = \left( \sum_{t=1}^{T} \alpha_t h_t(x) \right) \dots \quad (3.8)$$

alpha_t = 0.5 * ln((1 — e)/e)

For input x output of weak classifier t is H(x).

The assigned weight (alpha t) to the classifier can be determined as follows:

Weight of classifier's is simple; it can be determined by the error rate e.

**Silhouette Coefficient**

We used the silhouette coefficient to assess the efficacy of the clustering methods. The value of the silhouette score is between -1 and 1. 1 indicates that clusters are easily distinguishable and widely separated. The value "0" indicates that clusters are not distinguishable and that the distance between clusters is unimportant. '-1' denotes improper cluster assignment. In **Figure 4** we can see Sillhouette plot for training dataset, wherein three clusters are shown for training dataset which achieved the most prominent Silhouette score.

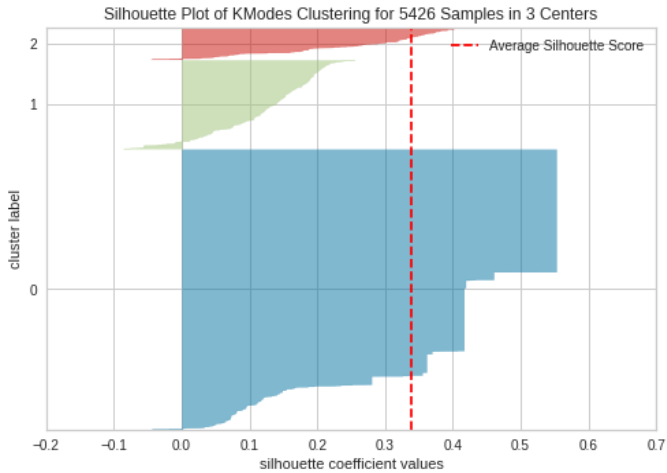

*Figure 4 Silhouette plot of KModes clustering*

**Distortion Score Elbow**

Using the elbow strategy, the ideal number of clusters was determined. Changing the value of k displays the value of the cost function graphically. As 'k' grows, the mean distortion will reduce, every cluster will include fewer instances, and instances will be situated closer to their specific centroids. Furthermore, as 'k' increases, the mean distortion reduction will become less rapid. We stopped further clustering the data at the elbow, which is the value of k at which the improvement in distortion drops to its maximum. As a result, three clusters have been suggested using the right elbow technique for MACCSkeys features (shown in **Figure 5**) and two clusters have been suggested for physical features including first cluster feature formed using MACCSkeys.

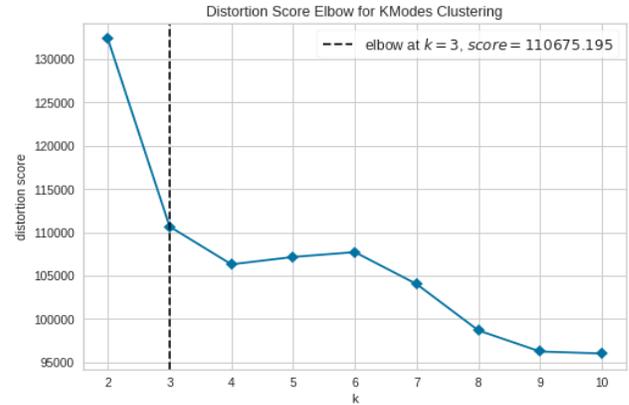

*Figure 5 Distortion Score Elbow plot for KModes Clustering*

### 3. RESULTS

**TABLE 1** *Results of all machine learning techniques for cluster dataset*

|  | Model | Accuracy | AUC | Recall | Prec. | F1 | Kappa | MCC | TT (Sec) |
|---|---|---|---|---|---|---|---|---|---|
| rf | Random Forest Classifier | 0.8860 | 0.9175 | 0.9015 | 0.9213 | 0.9112 | 0.7521 | 0.7526 | 0.0840 |
| dt | Decision Tree Classifier | 0.8860 | 0.9176 | 0.9015 | 0.9213 | 0.9112 | 0.7521 | 0.7526 | 0.0040 |
| ada | Ada Boost Classifier | 0.8868 | 0.9160 | 0.9015 | 0.9224 | 0.9118 | 0.7538 | 0.7544 | 0.0360 |
| et | Extra Trees Classifier | 0.8860 | 0.9176 | 0.9015 | 0.9213 | 0.9112 | 0.7521 | 0.7526 | 0.0670 |
| gbc | Gradient Boosting Classifier | 0.8860 | 0.9178 | 0.9015 | 0.9213 | 0.9112 | 0.7521 | 0.7526 | 0.0290 |
| knn | K Neighbors Classifier | 0.8717 | 0.9042 | 0.8963 | 0.9057 | 0.9004 | 0.7199 | 0.7215 | 0.0120 |
| svm | SVM - Linear Kernel | 0.8774 | 0.0000 | 0.9102 | 0.9038 | 0.9063 | 0.7285 | 0.7311 | 0.0050 |
| lightgbm | Light Gradient Boosting Machine | 0.8860 | 0.9178 | 0.9015 | 0.9213 | 0.9112 | 0.7521 | 0.7526 | 0.2200 |
| nb | Naive Bayes | 0.7833 | 0.9117 | 0.9508 | 0.7699 | 0.8507 | 0.4710 | 0.5084 | 0.0040 |
| lr | Logistic Regression | 0.8330 | 0.9117 | 0.9351 | 0.8298 | 0.8791 | 0.6119 | 0.6249 | 0.8040 |
| qda | Quadratic Discriminant Analysis | 0.7833 | 0.9117 | 0.9508 | 0.7699 | 0.8507 | 0.4710 | 0.5084 | 0.0050 |
| ridge | Ridge Classifier | 0.7833 | 0.0000 | 0.9508 | 0.7699 | 0.8507 | 0.4710 | 0.5084 | 0.0050 |
| dummy | Dummy Classifier | 0.6494 | 0.5000 | 1.0000 | 0.6494 | 0.7874 | 0.0000 | 0.0000 | 0.0100 |
| lda | Linear Discriminant Analysis | 0.7833 | 0.9117 | 0.9508 | 0.7699 | 0.8507 | 0.4710 | 0.5084 | 0.0050 |

## Ada-Boost accurately predicts pathway assignment of metabolites

We hypothesized that ML methods can be used to assign metabolites to pathways based on their structural features. Thus, in this study, we created a sizable dataset of 3110 pathways, including all diseases along with all metabolite instances identified or unidentified in curated metabolic pathway databases for humans to cluster and empirically test algorithms for pathway prediction. We generated 201 metabolite features using their SMILES and compared their information content against the gold standards. A wide range of machine learning (ML) techniques, including clustering algorithms like K-mode and K-prototype, have been used to make clusters of important features. Feature selection and ensemble approaches were also applied to feature data. Using the pathways dataset, we evaluated the ML approaches and their accuracies for pathway prediction. The ML algorithm Ada-Boost method achieved the highest accuracy of 88.68% and F- measure of 0.91 as shown in **Table 1** and **Figure 6**.

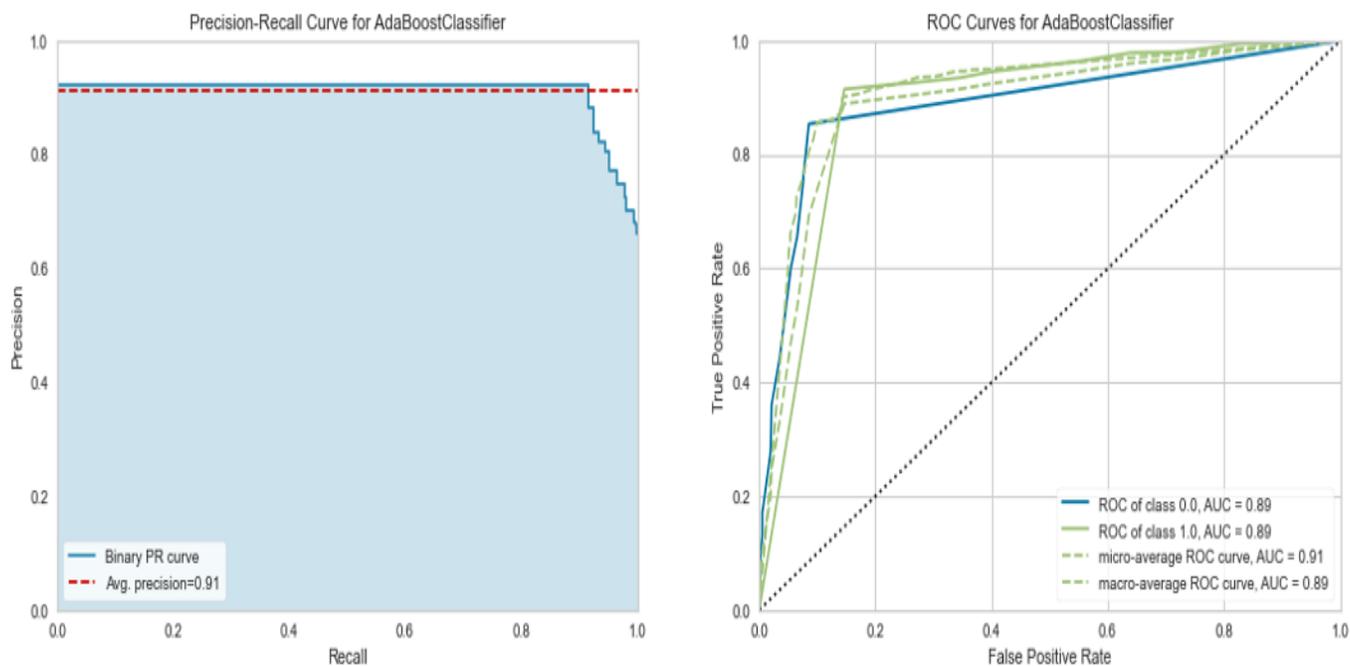

Figure 6 *ROC, AUC curve for AdaBoostClassifier and ROC and AUC curve for AdaBoostClassifier*

## Correlation Matrix Heatmap

With the success of boosting methods at assigning metabolites to pathways, we wondered which metabolite features were most informative. Thus, we quantified the correlation of all the features of the metabolites (**Figure 7**). Clusters formed using MACCSKeys have a 75% correlation, and physical features like "BondStereoCount" (Stereocenters or stereogenic atoms are the terms used to describe the carbon atoms that comprise the C=C double bond in 2-butene.) and "DeedBondStereoCount" have a 98% correlation with Pathways. Furthermore, "end_cluster" has a 92% correlation with pathways formed from all physical features in the training set. For the testing set, physical features like "BondStereoCount" and "DefinedBondStereoCount" have a correlation of approximately 80% with Pathways. As shown in **Figure 8**, "end_cluster" has a 67% correlation with pathways formed using all physical features in the training set, similar to the training set.

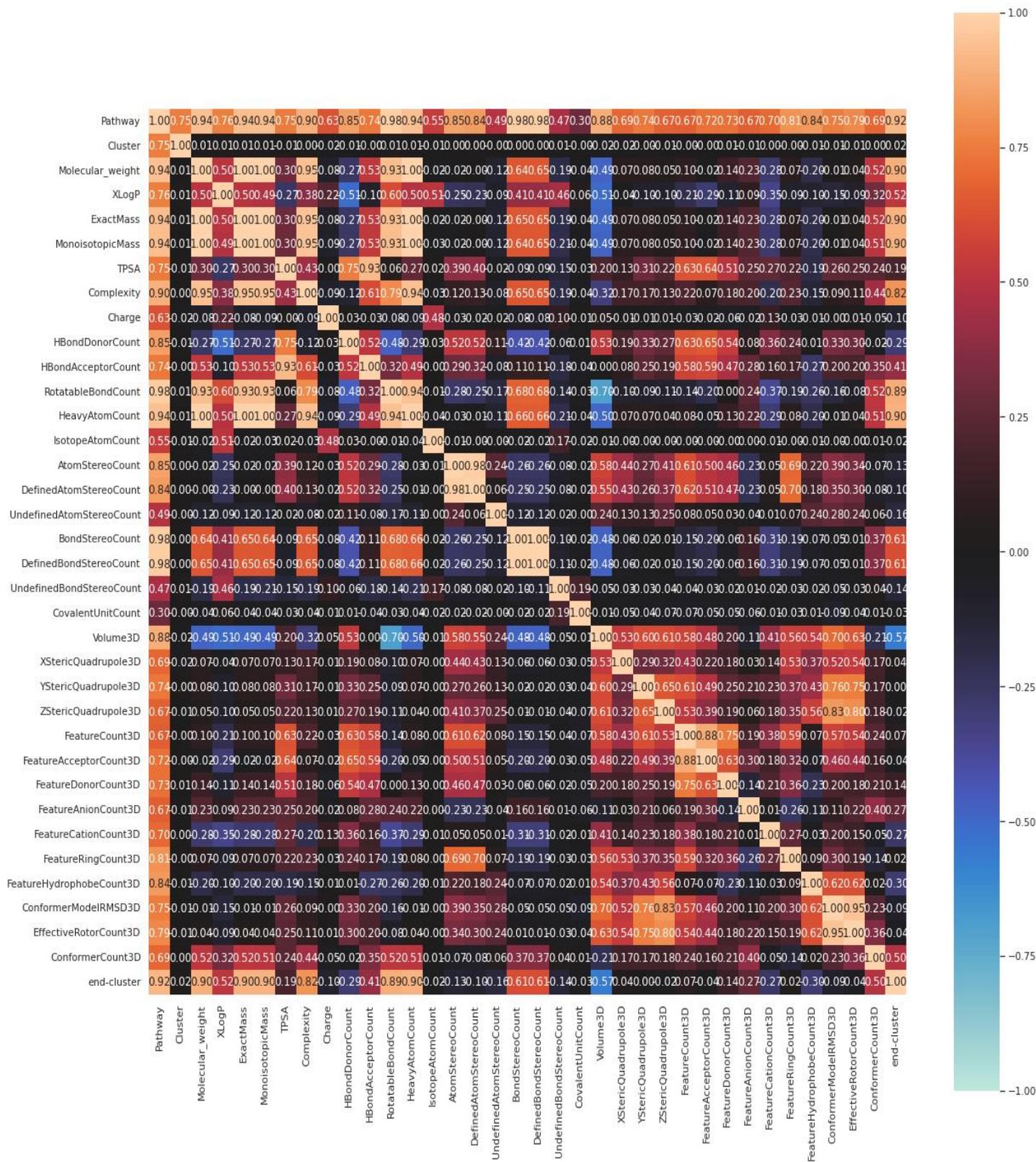

*Figure 7 Correlation heatmap for Training features*

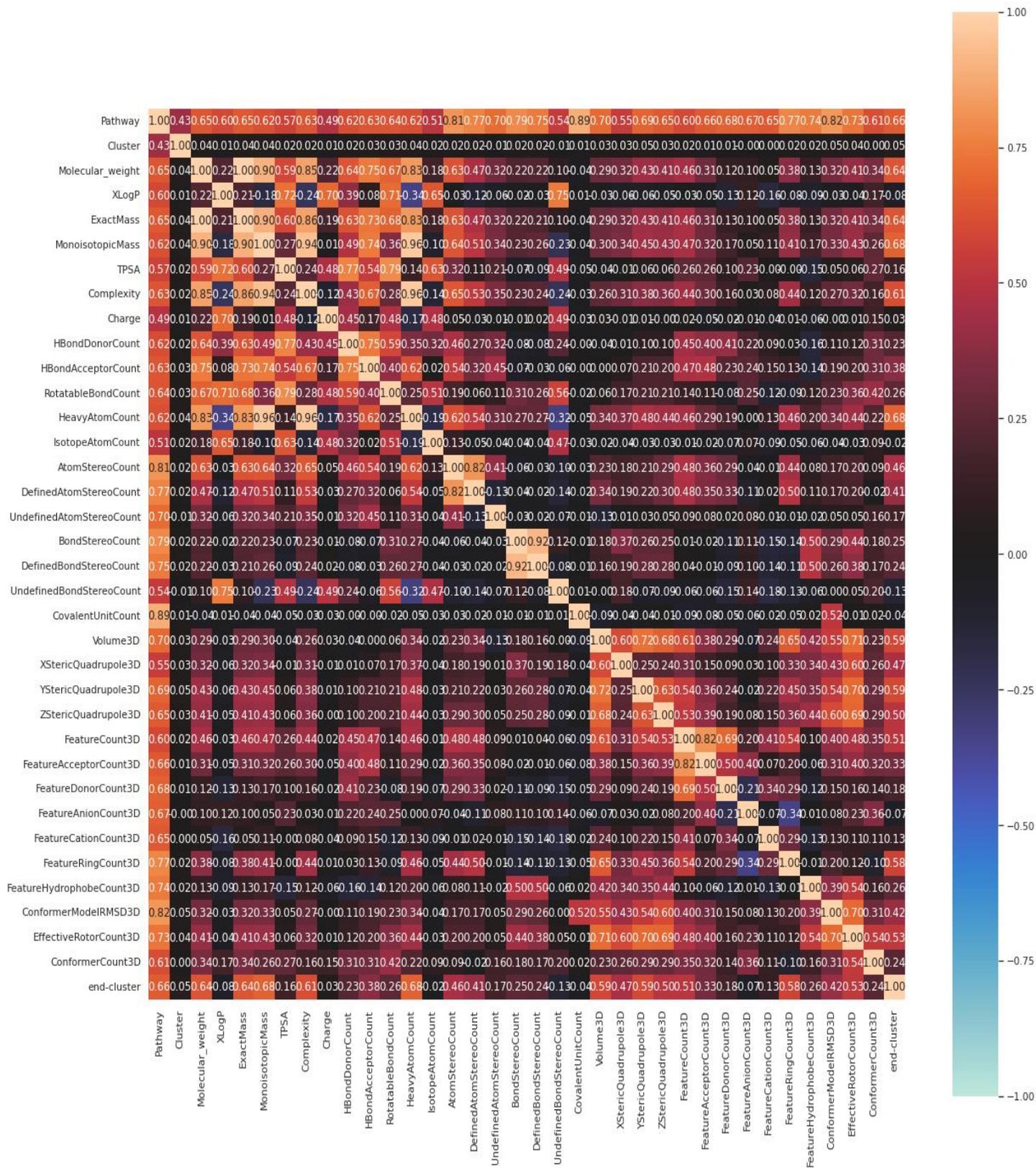

*Figure 8 Correlation heatmap for Testing features*

## 4. DISCUSSION

In this study, we evaluated classification accuracy using the prior models on the metabolite's dataset. The remaining 30% (including novel metabolites) were used as a testing set, while 70% were utilized as a training set.

Due to resonant structures and particular products that might be redundant and troublesome, reaction prediction continues to be a difficult challenge for studying metabolic pathways. Recent advances in machine learning have, however, improved performance (Cuperlovic-Culf, [2]). There are two different types of roadmaps for reaction prediction, depending on whether pairs of molecules or single molecules are included in the modelling: focusing compound pairs and focusing compounds.

For given chemicals, the techniques that focus on compounds first discover the products or precursors, then they generate the likely reactions. A substructure-based strategy, for instance, was described by Kotera et al. (2008) to find potential byproducts and/or antecedents of a given molecule and to produce a reasonable reaction. The target compound's structural relatives were sought out using RF techniques, in order to determine which of them would have the potential to become a product (or parent) of the target molecule in an enzyme-catalyzed reaction, the structural differences between them were next examined at. The same roadmap was followed by Wei et al. (2016). They initially constructed a neural network to predict the reaction type using a reaction fingerprinting technique from a given combination of reagents and reactants, after that, to estimate the probable product from reactants, they used SMARTS (Smiles Arbitrary Target Specification) conversion. Starting with reactant and reagent molecules, Depending on atom and bond descriptors, the neural network process counts all potential electron donors and deep dives inside the input molecules.

## 5. CONCLUSIONS

We demonstrated how metabolites can be used to predict metabolic pathways using machine learning techniques. The creation of a sizable dataset for pathway prediction, which we utilized to evaluate our algorithms, has been a major output of our effort. Our findings show that in bioinformatics, machine learning algorithms like K-mode and k-prototype can be exceptionally useful compared to other deep learning and clustering techniques for unsupervised data. We found that much of the information on whether a metabolite exists in a pathway or in multiple pathways is carried by a small number of features, i.e., molecular weight, complexity, HeavyAtomCount, BondStereoCount, etc. The most useful numerical metric was the physical properties of metabolites, which were used in the second group of features. As shown in **Figure 7** (confusion matrix), categorial features (MACCSkeys), which have clusters of all bits named as 'cluster', have a 75% correlation value with pathways and contributed as valuable features to the pathway prediction. While our current proposed method is not perfect, but with future work this approach would be combined with other techniques to potentially further enhance pathway predictions. Furthermore, increases in the amount of training data will also further enable increases in the accuracy of predictions. Combining new features with current features can make it simple to build and test the algorithms. It is also possible to test and implement new prediction algorithms with the currently available data. The machine learning methods we used give an estimate of the likelihood that a metabolite belongs to which cluster of pathways, rather than just categorizing and predicting the individual pathway. Moreover, researchers may also be able to find out if a metabolite belongs to a new pathway or not and can adjust the resultant pathway predictions to match their own choices for sensitivity vs. specificity.

## 6. REFERENCES

1. Wang L., Dash S., Ng C. Y., Maranas C. D. (2017). A Review of Computational Tools for Design and Reconstruction of Metabolic Pathways. Synth. Syst. Biotechnol. 2 (4), 243–252. 10.1016/j.synbio.2017.11.002 [Europe PMC free article] [Abstract] [CrossRef] [Google Scholar]

2. Cuperlovic-Culf M. (2018). Machine Learning Methods for Analysis of Metabolic Data and Metabolic Pathway Modeling, Metabolites, 8, 4. 10.3390/metabo8010004 [Europe PMC free article] [Abstract] [CrossRef] [Google Scholar]

3. Kim G. B., Kim W. J., Kim H. U., Lee S. Y. (2020). Machine Learning Applications in Systems Metabolic Engineering. Curr. Opin. Biotechnol. 64, 1–9. 10.1016/j.copbio.2019.08.010 [Abstract] [CrossRef] [Google Scholar]

4. Dale, J.M., Popescu, L. & Karp, P.D. Machine learning methods for metabolic pathway prediction. BMC Bioinformatics **11**, 15 (2010). https://doi.org/10.1186/1471-2105-11-15

5. Overbeek R, Begley T, Butler RM, Choudhuri JV, Chuang HY, Cohoon M, de Crecy-Lagard V, Diaz N, Disz T, Edwards R, Fonstein M, Frank ED, Gerdes S, Glass EM, Goesmann A, Hanson A, Iwata-Reuyl D, Jensen R, Jamshidi N, Krause L, Kubal M, Larsen N, Linke B, McHardy AC, Meyer F, Neuweger H, Olsen G, Olson R, Osterman A, Portnoy V, Pusch GD, Rodionov DA, Ruckert C, Steiner J, Stevens R, Thiele I, Vassieva O, Ye Y, Zagnitko O, Vonstein V. The subsystems approach to genome annotation and its use in the project to annotate 1000 genomes. Nuc Acids Res. 2005;**33**(17):5691–5702.
doi: 10.1093/nar/gki866. [PMC free article] [PubMed] [CrossRef] [Google Scholar]

6. DeJongh M, Formsma K, Boillot P, Gould J, Rycenga M, Best A. Toward the automated generation of genome-scale metabolic networks in the SEED. BMC Bioinformatics. 2007;**8**:139. doi: 10.1186/1471-2105-8-139. [PMC free article] [PubMed] [CrossRef] [Google Scholar]

7. Ye Y, Osterman A, Overbeek R, Godzik A. Automatic detection of subsystem/pathway variants in genome analysis. Bioinformatics. 2005;**21**(Suppl 1):i478–i486. doi: 10.1093/bioinformatics/bti1052. [PubMed] [CrossRef] [Google Scholar]

8. Matthews L, Gopinath G, Gillespie M, Caudy M, Croft D, de Bono B, Garapati P, Hemish J, Hermjakob H, Jassal B, Kanapin A, Lewis S, Mahajan S, May B, Schmidt E, Vastrik I, Wu G, Birney E, Stein L, D'Eustachio P. Reactome knowledgebase of human biological pathways and processes. Nuc Acids Res. 2009. pp. D619–22. [PMC free article] [PubMed] [CrossRef]

9. Okuda S, Yamada T, Hamajima M, Itoh M, Katayama T, Bork P, Goto S, Kanehisa M. KEGG Atlas mapping for global analysis of metabolic pathways. Nuc Acids Res. 2008;**36**:W423–26. doi: 10.1093/nar/gkn282. [PMC free article] [PubMed] [CrossRef] [Google Scholar]

10. Varma A, Palsson B. Metabolic Flux Balancing: Basic concepts, Scientific and Practical Use. Bio/Technology. 1994;**12**:994–8. doi: 10.1038/nbt1094-994. [CrossRef] [Google Scholar]

11. Liao L, Kim S, Tomb JF. Genome comparisons based on profiles of metabolic pathways. Proceedings of the 6th International Conference on Knowledge-Based Intelligent Information and Engineering Systems (KES 02) 2002. pp. 469–476.

12. Kastenmuller G, Gasteiger J, Mewes HW. An environmental perspective on large-scale genome clustering based on metabolic capabilities. Bioinformatics. 2008;**24**(16):i56–62. doi: 10.1093/bioinformatics/btn302. [PubMed] [CrossRef] [Google Scholar]

13. Kastenmuller G, Schenk ME, Gasteiger J, Mewes HW. Uncovering metabolic pathways relevant to phenotypic traits of microbial genomes. Genome Biol. 2009;**10**(3):R28. doi: 10.1186/gb-2009-10-3-r28. [PMC free article] [PubMed] [CrossRef] [Google Scholar]

14. Sun J, Zeng AP. IdentiCS - Identification of coding sequence and in silico reconstruction of the metabolic network directly from unannotated low-coverage bacterial genome sequence. BMC Bioinformatics. 2004;**5**:112. doi: 10.1186/1471-2105-5-112. [PMC free article] [PubMed] [CrossRef] [Google Scholar]

15. Pinney JW, Shirley MW, McConkey GA, Westhead DR. metaSHARK: software for automated metabolic network prediction from DNA sequence and its application to the genomes of Plasmodium falciparum and Eimeria tenella. Nucleic Acids Research. 2005;**33**(4):1399–1409. doi: 10.1093/nar/gki285. [PMC free article] [PubMed] [CrossRef] [Google Scholar]

16. Pireddu L, Poulin B, Szafron D, Lu P, Wishart DS. Pathway Analyst -- Automated Metabolic Pathway Prediction. Computational Intelligence in Bioinformatics and Computational Biology, 2005. CIBCB '05. Proceedings of the 2005 IEEE Symposium on. 2005. pp. 1–8. full_text.

17. Pireddu L, Szafron D, Lu P, Greiner R. The Path-A metabolic pathway prediction web server. Nucleic Acids Research. 2006;**34**(suppl 2):W714–719. doi: 10.1093/nar/gkl228. [PMC free article] [PubMed] [CrossRef] [Google Scholar]

18. McShan D, Rao S, Shah I. PathMiner: Predicting metabolic pathways by heuristic search. Bioinformatics. 2003;**19**(13):1692–8. doi: 10.1093/bioinformatics/btg217. [PMC free article] [PubMed] [CrossRef] [Google Scholar]

19. Cakmak A, Ozsoyoglu G. Mining biological networks for unknown pathways. Bioinformatics. 2007;**23**(20):2775–2783. doi: 10.1093/bioinformatics/btm409. [PubMed] [CrossRef] [Google Scholar]

20. Yamanishi Y, Vert JP, Kanehisa M. Supervised enzyme network inference from the integration of genomic data and chemical information. Bioinformatics. 2005;**21**(suppl 1):i468–477. doi: 10.1093/bioinformatics/bti1012. [PubMed] [CrossRef] [Google Scholar]

21. Huang Z. Extensions to the k – Means algorithm for Clustering Large Data Sets with Categorical Values. Data Mining and Knowledge Discovery, 1988 , 2: 283 – 304

22. Aranganayagi S, Thangavel K,Sujatha S. New Distance Measure based on the Domain for Categorical Data.

23. Huang Z. 1997a. Clustering large data sets with mixed numeric and categorical values. Proceedings of the First Pacific Asia Knowledge Discovery and Data Mining Conference, Singapore: World Scientific, pp. 21–34.

24. Huang Z. 1997b. A fast clustering algorithm to cluster very large categorical data sets in data mining.

*Proceedings of the SIGMOD Workshop on Research Issues on Data Mining and Knowledge Discovery, Dept. of Computer Science, The University of British Columbia, Canada, pp. 1–8.*

25. *Han J, Kamber M. Data Mining Concepts and Techniques, Morgan Kaufmann, San Francisco, 2001.*

26. *Hsu C.C, Huang Y.P, Incremental clustering of mixed data based on distance hierarchy, Expert Systems with Applications 35 (3) (2008) 1177–1185.*